\documentclass[preprint,12pt]{elsarticle}

\usepackage{graphicx}
\usepackage{mathptmx}
\usepackage{amsmath}

\usepackage{amssymb}

\begin{document}

\begin{frontmatter}



\title{Firing statistics of inhibitory neuron with delayed feedback. I. Output ISI
probability density}


\author{A. K. Vidybida}
\ead{vidybida@bitp.kiev.ua}
\ead[url]{http://www.bitp.kiev.ua/pers/vidybida}
\author{K. G. Kravchuk}
\address{Bogolyubov Institute for Theoretical Physics, 
Metrologichna str. 14-B, 03680 Kyiv, Ukraine}
\ead{kgkravchuk@bitp.kiev.ua}

\begin{abstract}
Activity of inhibitory neuron with delayed feedback is considered
in the framework of point stochastic processes. The neuron receives excitatory input impulses from a Poisson stream, and inhibitory impulses from the feedback line with a delay. We investigate here, how does the presence of inhibitory feedback  affect the output firing statistics.
Using binding neuron (BN) as a model, we derive analytically the exact expressions for
the output interspike intervals (ISI) probability density, mean output ISI and coefficient of variation
as functions of model's parameters  for the case of threshold 2.
Using the leaky integrate-and-fire (LIF) model, as well as the
BN model with higher thresholds, 
these statistical quantities are found numerically. In contrast to the
previously studied situation of no feedback, the
ISI probability densities found here both for BN and LIF neuron 
become bimodal and have discontinuity of jump type.
Nevertheless, 
the presence of inhibitory delayed feedback was not found
to affect substantially the output ISI coefficient of variation.
The ISI coefficient of variation found ranges between 0.5 and 1. 
It is concluded that introduction of delayed inhibitory feedback 
can radically change neuronal output firing statistics. This statistics is as well
distinct from what was found previously, \cite{BNDF}, by a similar
method for excitatory neuron with delayed feedback.
\end{abstract}

\begin{keyword}
Inhibitory neuron \sep Delayed feedback \sep Poisson process \sep 
Interspike intervals probability density \sep Coefficient of variation
\end{keyword}
\end{frontmatter}


\section{Introduction}\label{intro}

A realistic neuronal network is normally characterized with a complicated system of 
excitatory and inhibitory interconnections between individual neurons the network
is composed of. Statistics of spiking activity of individual neurons can be measured
experimentally \cite{Nawrot2007,Segundo1963,Softky}. It would be interesting to understand how the details of 
network's construction might influence statistics of neuronal activity, when the 
network is driven with some stimulation, or allowed reverberating freely. Exact 
theoretical analysis of this question in a developed network represents fair 
mathematical difficulties. 
At the same time, numerous studies suggest that feedback and delays in
the intercomponent communication can be essential
factors in determining activity of a composed system, see, 
e.g. \cite{Kostur,Adeli,Zhang}.

In a real neural network, 
constructing elements can be found, which allow exact mathematical treatment.
The results of such a treatment can shed light on the nature of transformations
the neuronal activity might undergo while spreading within a real neural network.
One example is an excitatory neuron which sends its output impulses onto its own
dendritic tree --- the neuron with excitatory feedback. This type of constructive
element has been found in the olfactory bulb \cite{Aron,Nicoll}.
Theoretical study of this construction fed with Poisson stream revealed interesting 
peculiarities in its output activity statistics \cite{BNDF}. 
Another natural variant of this
construction is a neuron with inhibitory feedback. It seems that selfinhibition happens
more frequently in the brain, than selfexcitation. Selfinhibition can be
slow, due to potassium channels opening \cite{Bacci2004}, or fast, due to chlorine channels \cite{Bacci,Sakmann}. It also can be direct (through autapses) \cite{Bacci,Nicoll,Smith}, or acting through a single intermediate neuron
\cite{Sakmann}. Also, it can be evoked not only by means of a spike delivered to corresponding synaptic connection, but also through
extended diffusion of some usual \cite{Smith}, or unusual \cite{Bacci2004} mediator. 

In this paper, we consider situation of inhibitory neuron fed externally
with excitatory impulses from the Poisson stream. 
The neuron sends its output impulses to its own
input through feedback line with delay. Both input and output streams are
treated as point stochastic processes with no diffusion approximation applied.
Our purpose is to find the probability density function (PDF) of the output 
interspike intervals and to study its properties, as well as to compare those 
quantities for two neuronal models, namely the binding neuron and the leaky
integrate and fire neuron.

\section{Methods}
\subsection{BN without feedback}
\label{BNdef}

The binding neuron model \cite{Vid3} is inspired by numerical simulation \cite{Vid} of Hodg\-kin-Hux\-ley-type point neuron, as well as by the leaky in\-teg\-rate-and-fire (LIF) model \cite{Segundo}. In the binding neuron, the trace of an input is 
remembered for a fixed period of time after which it disappears completely. 
This is in the contrast with the above two models, where the postsynaptic potentials decay exponentially and can be forgotten only after triggering. 
The finiteness of memory in the binding neuron allows one to obtain
exact mathematical conclusions concerning its firing statistics beyond
the diffusion approximation technique. 
Recently, the finiteness is utilized for exact mathematical description of the output sto\-c\-h\-astic process if the binding neuron is driven with the Poisson 
input stream in the case of no feedback, \cite{Vid5}, 
for BN with instantaneous feedback, \cite{BNF} and for BN with delayed excitatory feedback, \cite{BNDF}. 

\begin{figure} [h]
\unitlength=0.9mm
\begin{center}
	\includegraphics[width=0.69\textwidth]{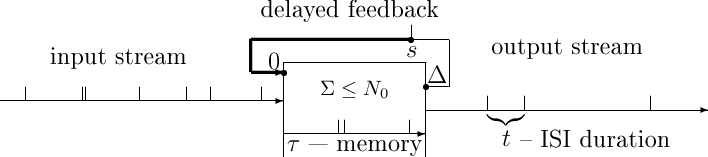}
\end{center}
\caption{\label{BNF} Binding neuron with feedback (see \cite{Vid3} for details). $\tau$ is similar to the ``tolerance interval'' discussed in \cite[p. 42]{MacKay}. Multiple input lines with Poisson streams are joined into a single one here.}
\end{figure}

The BN works as follows (see Fig. \ref{BNF} with the feedback line removed). All excitatory input impulses have the same magnitude. Each one of them is stored in the BN for a fixed period of time, $\tau$, and then it is forgotten. When the number of stored excitatory impulses, $\Sigma$, becomes equal to the BN's threshold, $N_0$, the BN fires an output spike, clears its internal memory, 
and is ready to receive fresh inputs. 
Thus, the state just after firing corresponds to the resting state of
excitable membrane in real neurons, and the presence of impulses in the internal
memory of BN corresponds to partially depolarized state.
In this work, we take BN with $N_0=2$ for analytic derivation. BNs with higher thresholds are studied numerically in Sec.~\ref{num}.  

Normally, any neuron has a number of input lines. If input stream in each line is Poissonian and all lines have the same weight, all of them can be joined into a single one, like in Fig.~\ref{BNF}, with intensity, $\lambda$, equal to sum of intensities in the individual lines. 

The output statistics for BN with  $N_0=2$ and without feedback was studied 
before. Here, we will need 
the ISI probability density function for BN without feedback, 
$P^0(t)$, where $t>0$ denotes the output ISI duration, which was 
obtained in \cite[Eq. (3)]{Vid5} as
\begin{equation}
\label{P0}
	m\tau\le t\le (m+1)\tau
	 \,\Rightarrow\,
	 P^{0}(t) = y_{m} (t),\,
	m=0,1,\dots,
\end{equation}
where the functions $y_{m}(t)$ are defined according to the following 
recurrent relation:
\begin{equation}\label{y0}
	y_{0}(t) = e^{-\lambda t} \lambda^{2}\ t,
\end{equation}
\begin{multline}\label{ym}
y_{m+1}(t) =y_{m} (t) + \frac{\lambda^{m+3}}{(m+2)!}\ (t-(m+1)\tau)^{m+2} e^{-\lambda t} 
\allowdisplaybreaks\\[0.5\baselineskip]
		-\frac{\lambda^{m+2}}{(m+1)!}\ (t-(m+1)\tau)^{m+1} e^{-\lambda t},
	\qquad m=0,1,\ldots.
\end{multline}
$P^0(t)$ is a uni-modal function,
which reaches its maximum at $t=\min(1/\lambda;\tau)$.

The first moment, $W_1^0$, of the probability density (\ref{P0}) was found
in \cite{Vid5} as
\begin{equation}
\label{W10}
	W_1^0\equiv\int_{0}^{\infty} t\,P^0(t)\,dt=
	\frac{1}{\lambda}\left(2+{1\over e^{\lambda\tau}-1}\right),
\end{equation}
which will be used later.
Further, we also utilize the probability $\Pi(t)$ to get from BN without feedback
an output ISI, which is longer than $t$:
\begin{equation}
\label{Pidef}\nonumber
	\Pi(t)\equiv \int_{t}^{\infty}P^{0}(t')\ dt'=1-\int_{0}^{t}P^{0}(t')\ dt'.
\end{equation}
We will need the expression for $\Pi(t)$ only in the domain $0\le t\le\tau$, where it can be found by substituting $y_0(t')$, (\ref{y0}), instead of $P^0(t')$ above:
\begin{equation}
\label{Pi}
	\Pi(t) = (1+ \lambda t )\ e^{- \lambda t },
	\qquad 0\le t\le\tau.
\end{equation}

\subsection{Feedback line action}

In this work, we consider the situation, when BN receives excitatory input from the Poisson stream and inhibitory impulses from the feedback line.

We assume, that time delay $\Delta$ of impulse in the feedback line is 
fixed and is smaller than the BN's memory duration, $\tau$:
\begin{equation}\label{Dlt}
\Delta<\tau.
\end{equation}
It allows us to make analytic expressions shorter\footnote{We were 
unable to found
experimental value for $\Delta$ in an autaptic connection. A crude
estimate can be made based on the action potential propagation velocity,
$v$, and the connection length, $l$. We put $v=0.5$ m/s (the smallest known).
Expect that the autaptic connection is confined within a cortical minicolumn.
The minicolumn diameter is about 50 $\mu$m. This gives
for the propagation delay $\Delta_p\sim 3\cdot l/v = 0.3$ ms. 
In this case, the delay of the feedback communication, $\Delta$, is mainly
due to the EPSP rise time. The rise time usually falls between 1 ms and 
10 ms, which to an extent supports (\ref{Dlt})  if one takes 
into account that $\tau$ should be comparable with the membrane
relaxation constant, $\tau_M$.
If one expects that the connection is confined within a cortical hypercolumn,
then $\Delta_p\sim 3$ ms, which still does not refute (\ref{Dlt}). 
But if an autaptic connection is considerably longer, then 
condition, which is opposite to (\ref{Dlt}) may be correct.}. 

If the line is empty, when neuron fires, an output impulse enters the line and after delay $\Delta$ reaches neuron's input. If the line already keeps an impulse
at the moments of BN firing, it does not accept a new one. 
It means, that at any given moment, the feedback line either conveys 
one impulse, or it is empty.
The state of the nonempty feedback line can be described with a single number,
$s$, $s\in]0;\Delta]$,
which gives the time to live of the impulse in the feedback line.
The time to live determines position of the impulse in the feedback line, 
see Fig~\ref{BNF}, thick part of the feedback line. 
It equals to the time, which is required for the impulse to reach
the end of the feedback line, if starting from a given position.
The values of $s$ are used just at the moments of output ISI beginnings
(just after BN firings). At these moments, the feedback line is never empty.

The inhibitory action of feedback impulses is modelled in the following way. 
When the inhibitory impulse reaches BN, it annihilates all excitatory impulses already present in the BN's memory, similarly as Cl-type inhibition shunts 
depolarization of excitable membrane, see \cite{Sch}.
If at the moment of inhibitory impulse arrival, the BN is empty, then
the impulse disappears without any action, similarly as Cl-type inhibition
does not affect membrane's voltage in its resting state.
Such inhibition is "fast" in that sense, that the inhibitory impulses act instantaneously and are not remembered by neuron.

\subsection{Derivation outline}
\label{outline}

It is clear, that both the binding neuron and the feedback line operate in deterministic manner.
Nevertheless, the probabilistic description is required for the output stream 
because of the stochastic nature of the driving Poisson process. 

Let us denote by $P^\Delta(t)$ the ISI probability density function
for neuron with delayed inhibitory feedback.
In order to calculate $P^\Delta(t)$, we use the procedure,
previously utilized for BN with excitatory delayed feedback \cite{BNDF}.
For this purpose we define an auxiliary random variable $S$, wich values
$s\in]0;\Delta]$ are the values of time to live of the impulse in the feedback line
at the beginning of an interspike interval. A possibility to introduce such a variable
is ensured by the fact that, at the beginning of any ISI, the feedback line is never empty,
see pervious section. It is clear that both $S$ and $T$ outcomes are uniquely
determined by the unique outcome of the driving Poisson process. Here $T$ denotes the
ISI random variable. The difference between $T$ and $S$ is that the $S$ outcome becomes
known in physical time before the corresponding outcome of $T$ is known. That is,
if at the beginning of an ISI, the $S$ outcome is $s$, then this $s$ together with the 
consequent outcome of the driving Poisson process determine the outcome $t$,
corresponding to this $s$. Due to this fact it is natural to define 
the conditional probability 
density, $P^\Delta(t\mid s)$. Namely, $P^\Delta(t\mid s)\,dt$ gives the probability to obtain an output ISI of 
duration within interval
$[t;t+dt[$, provided there was an impulse in the feedback line with time to 
live equal $s$ at the moment of this ISI beginning.

As the first step, we  calculate 
the conditional probability density, $P^\Delta(t\mid s)$.

Then, we calculate the probability density, $f(s)$, $s\in]0;\Delta]$
for the variable $S$.

The output ISI probability density can be calculated based on the expressions for $P^{\Delta}(t\mid s)$ and $f(s)$, namely:
\begin{equation}
\label{usered}
	P^\Delta(t) = \int_{0}^{\Delta} P^\Delta(t\mid s) f(s)\,ds.
\end{equation}

In order to find $f(s)$, we first obtain the transition probability density $P(s'\mid s),\ s,s'\in ]0;\Delta]$, which gives the probability that at the beginning of some output ISI, the line has an impulse with time to live within the interval $[s';s'+ds'[$, provided that at the beginning of the previous ISI it had impulse with time to live equal $s$. 
$f(s)$ is then found as normalized to 1 solution of the following equation:
\begin{equation}
\label{umoriv}
	\int_{0}^{\Delta} P(s'\mid s)\,f(s)\,ds=f(s').
\end{equation}

\section{Main calculation}
\label{main}
\subsection{Conditional probability density $P^{\Delta}(t\mid s)$}

In order to derive $P^\Delta(t\mid s)$, domains $t<s$ and $t\ge s$ should be considered separately.

In the case $t<s$, the output impulse must be generated without the line impulse involved. Therefore, probability density for such ISI values is the same as for BN without any feedback:
\begin{equation}
\label{vyp1}
	P^\Delta(t\mid s)=P^0(t),\quad t<s.
\end{equation}
Here $P^0(t)$ is the output ISI probability density for BN without feedback, given in Eq.~(\ref{P0}).

At the moment $t=s$, the inhibitory feedback impulse reaches the BN and BN 
becomes empty. To trigger the neuron within the infinitesimal interval 
$[s;s+dt[$, one needs
to get two input impulses within this interval. Probability of this event is of order
$dt^2$. Therefore, $P^{\Delta}(t\mid s)=0$ at $t=s$.

In order to obtain ISI $t> s$, two independent events must occur: (i) BN without feedback fires no spikes during time interval $]0;s]$; (ii) BN without feedback starts empty at moment $s$ and is firstly triggered at moment $t$. These events are independent since their realizations are defined by behavior of Poisson input stream on disjoint intervals $]0;s]$ and $]s;t]$.  By definition of $\Pi(t)$, see Eq.~(\ref{Pidef}), the probability to have (i) is $\Pi(s)$, and (ii) has the probability $P^0(t-s)\,dt$. Therefore,
\begin{equation}
\label{rozpdt}
	P^\Delta(t\mid s)= \Pi(s)\ P^0(t-s),
	\quad t> s.
\end{equation}
Taking into account Eq.~(\ref{Pi}), (\ref{vyp1})  and (\ref{rozpdt}) for the case $\Delta<\tau$ one obtains $P^{\Delta}(t\mid s)$ as follows:
\begin{equation}
\label{rozpdr0}
	P^{\Delta }(t\mid s)=
	\begin{cases}
		\lambda^2 t\ e^{-\lambda t},\quad t\in]0;s[,\\\\
		(1+\lambda s)\ e^{-\lambda s}P^0(t-s),\quad t\ge s.\quad 
	\end{cases}
\end{equation}

The conditional probability density $P^{\Delta}(t\mid s)$, given in (\ref{rozpdr0}),
is normalized: {\small $\int_{0}^{\infty} P^{\Delta}(t\mid s)\,dt=1$}.
Also, $P^{\Delta }(t\mid s)$ has a jump discontinuity
of height $\lambda^2 s\ e^{-\lambda s}$ at $t=s$.

\subsection{Transition probability density $P(s'\mid s)$}

From the definition of the transition probability $P(s'\mid s)$ given in the last
paragraph of Sec. \ref{outline} it follows that
$$
s'\ge s\quad \mathbf{and}\quad s'\ne\Delta\quad \Rightarrow\quad P(s'\mid s)=0.
$$
Indeed, consider the pair $(t,s)$, wher $t$ is the ISI duration and $s$ is
the impulse time to live in the feedback line when this ISI starts.
If for this pair the inequality $t\le s$ takes place, then $s'=s-t$
and $s'<s$ with necessity. In the opposite situation when $t>s$,
the line becomes empty before the end of the ISI $t$. At the end of $t$,
the neuron starts next ISI by firing a spike, which charges the line
with a fresh impulse. This means that in the next pair $(t',s')$,
with necessity, $s'=\Delta$.
Therefore, the set of values $(s',s)$, 
where $P(s'\mid s)$ still has to be found is defined
by the following relations:
$$
s'<s\quad\mathbf{or}\quad s'=\Delta.
$$
From the meaning of $P^\Delta(t\mid s)$ it follows that Eq.~(\ref{vyp1}) allows 
one to calculate $P(s'\mid s)$ for $s'<s$, namely:
\begin{equation}
\label{umov0}
	P(s'\mid s) = P^{\Delta}(s-s' \mid s) = P^0(s-s')
	=e^{-\lambda(s-s')}\lambda^2(s-s'),
	\quad s'< s\in ]0;\Delta],
\end{equation}
where (\ref{P0}), (\ref{y0}) and (\ref{Dlt}) were used.

Consider the exact equality $s'=\Delta$. It is fulfilled every time, when for previous
ISI the inequality $t\ge s$ holds, and this inequality happens with non-zero
probability. Therefore, the probability density $P(s'\mid s)$ has singularity of $\delta$-function type at $s'=\Delta$. For calculating its mass  it is enough to utilize
the normalization condition:
$$
\int\limits_0^\Delta P(s'\mid s) ds' =1,\quad s\in]0;\Delta],
$$
which gives
\begin{equation}
\label{umov2}
	P(s'\mid s) = 
	\begin{cases}
		e^{-\lambda(s-s')}\lambda^2(s-s'), \qquad s' < s\in ]0;\Delta],\\
		\left(\lambda\,s+1\right)\, e^ {- \lambda\,s }\ \delta(s'-\Delta), \quad s'\ge s\in ]0;\Delta].
	\end{cases}
\end{equation}

\subsection{Delays probability density}

In order to find $f(s)$,
one should substitute $P(s'\mid s)$ from (\ref{umov2})
into (\ref{umoriv})
and solve the obtained equation.
As $P(s'\mid s)$ obtained here is exactly the same as for BN with excitatory 
delayed feedback \cite[Eq. (9)]{BNDF},
the equation for $f(s)$ and $f(s)$ itself will be the same also.
In \cite[Eq. (10)]{BNDF}, the probability density $f(s)$ was obtained as
\begin{equation}
\label{vygliad}
	f(s)=a\,\delta(s-\Delta)+g(s),
\end{equation}
where $g(s)$ -- is an ordinary function, which vanishes out of interval $]0;\Delta]$: 
\begin{equation}
\label{groz}
	g(s)=\frac{a\,\lambda}{2}\left(1-e^{- 2\lambda(\Delta-s)}\right),
	\quad s\in]0;\Delta],
\end{equation}
and $a$ -- is the dimensionless constant:
\begin{equation}
\label{a}
a=4e^{2\lambda\Delta}/\left((2\lambda\Delta+3)e^{2\lambda\Delta}+1\right),
\end{equation}
which gives the probability to find the impulse in the feedback line with time to live $\Delta$ at the beginning of any ISI.

\subsection{Numerical simulations}\label{MeNum}

Numerical simulations were carried out here for several purposes. 
The first one was to check numerically correctness of the expressions found 
analytically.
The second one was to obtain ISI distributions for 
higher thresholds and for threshold 2 with $\Delta>\tau$.
The third one was to compare the ISI distributions found 
here for the binding neuron model with those for the leaky integrate and fire (LIF) 
model.
 
 A C$^{++}$ program was developed, which allows us to calculate all the necessary probability distributions.  The program includes the BNDF class, which 
 analyzes the input stream and fires in accordance with the rules, 
 described above. The Poisson steams of various intensity were produced with 
 the help of the 
 GNU Scientific Library\footnote{see http://www.gnu.org/software/gsl/}. 
 With the help of our program, 
 output stream samples were produced by calculating up to $N=30\,000\,000$ 
 output spikes.  The samples were scanned for interspike intervals of various 
 duration, and the probability density distribution was then calculated by 
 normalization. 

The LIF neuron was simulated in its simplest version. Namely, the neuron's 
state at any moment of time $\vartheta$ is completely characterized by its 
membrane voltage at that moment, $V(\vartheta)$. Without stimulation,
the $V(\vartheta)$ decays exponentially to the resting state with $V=0$:
$$
V(\vartheta+t)=e^{-t/\tau_M}\,V(\vartheta),
$$
where $\tau_M$ -- is the membrane relaxation time.
An input impulse advances $V$ by a fixed value, $y_0$, instantaneously:
$$V\quad\to\quad V+y_0,$$
where $y_0$ mimics the EPSP peak value. If the resulting voltage satisfies the
inequality
$$V+y_0>V_0,$$
where $V_0$ -- is the firing threshold, then the LIF neuron fires an output spike
and appears in the resting state.

For numerical simulations we choose $\tau_M=10$ ms, $y_0=4$ mV, 
$V_0=5$ mV. %
These values are comparable with those found in the inhibitory interneurons
of CA3 hippocampal region, \cite{Miles}. 
The relation between $V_0$ and $y_0$ ensures that two input impulses are
able to trigger the LIF provided they are close in time. For the  
inhibitory interneurons, this is because of their
depolarized resting state, \cite{Jonas}.
It is reported, \cite{Gulyas}, that even single impulse from a piramidal cell 
may trigger interneuron of this type. Interesting, that selfinhibition is
found in the inhibitory interneurons also, but in the neocortex, \cite{Bacci}.

\section{Results}
\subsection{ISI probability density}
\label{prob}
%
%
%
%
%
%
\begin{figure}
	\includegraphics[angle=-90,width=0.5\textwidth]{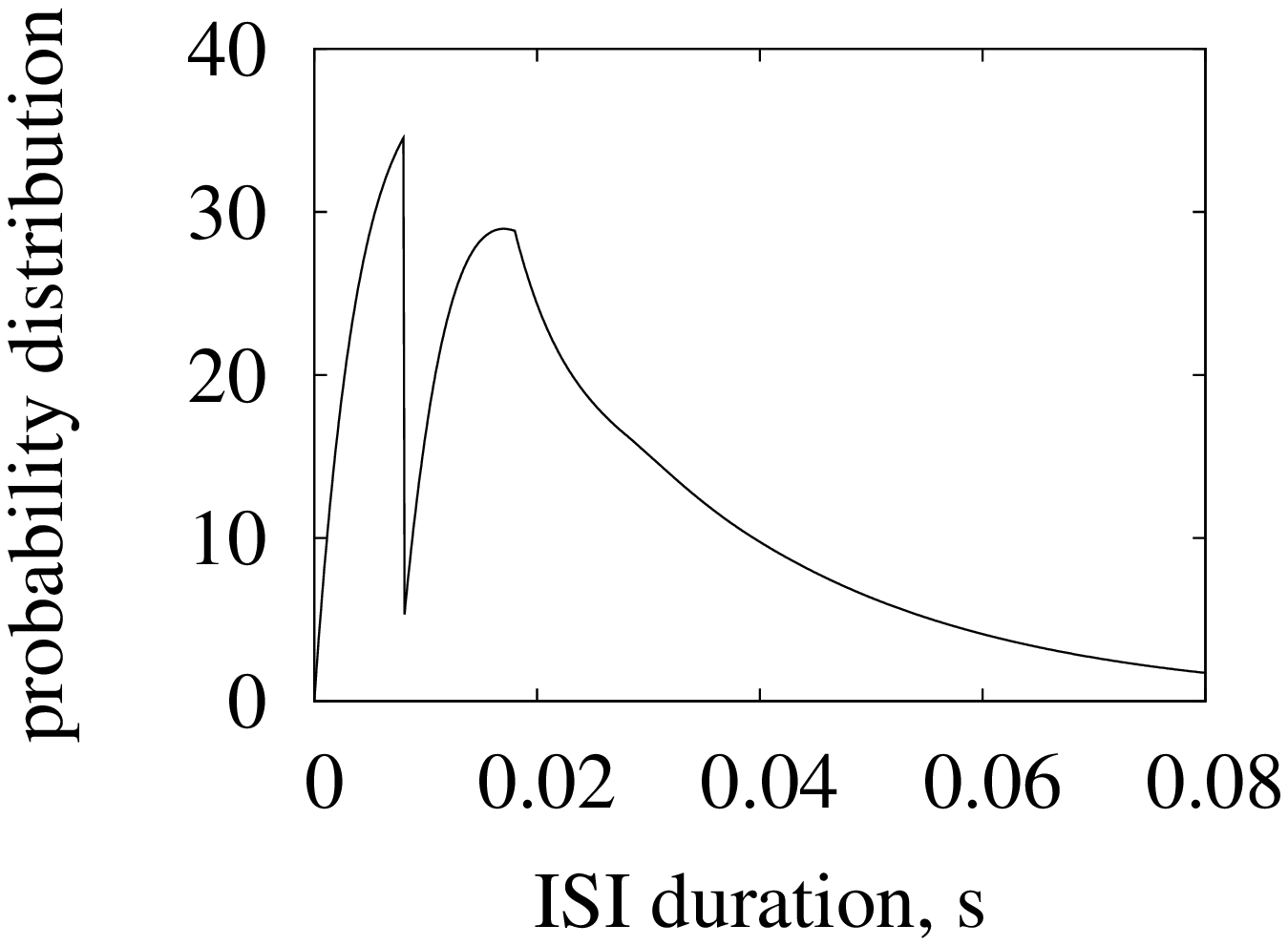}%
	\includegraphics[angle=-90,width=0.5\textwidth]{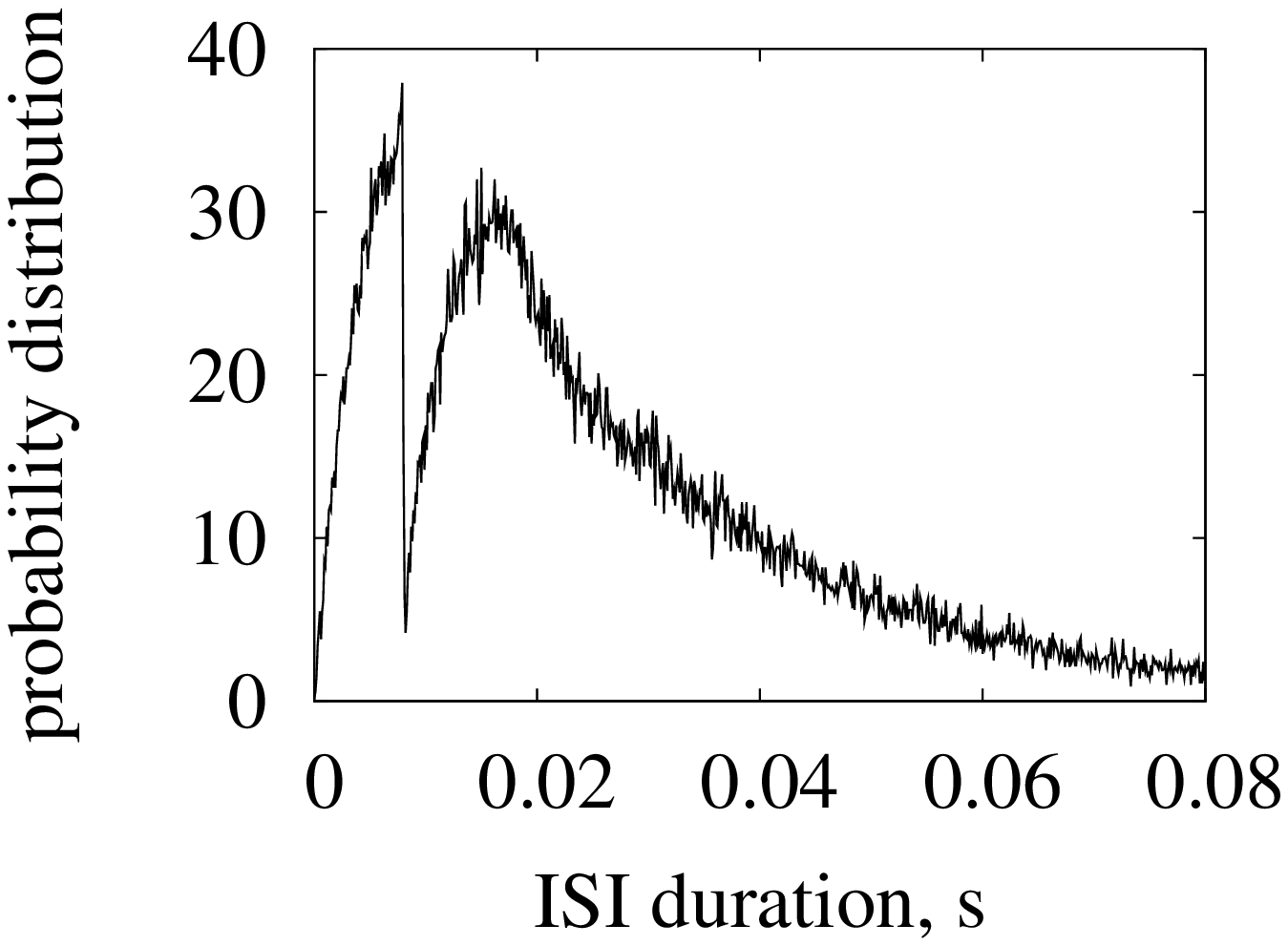}%
\caption{\label{ISI}%
Example of ISI PDF. \emph{Left} --- calculation in accordance
with Eqs. (\ref{t-A}), (\ref{eq:P1}), (\ref{eq:P2}). \emph{Right} --- numerical 
simulation. 
For both panels:  $\tau=10$ ms, $\Delta=8$ ms, $\lambda=10$ s$^{-1}$, 
$N_0=2$.}
\end{figure}

\begin{figure}
	\includegraphics[angle=-90,width=0.45\textwidth]{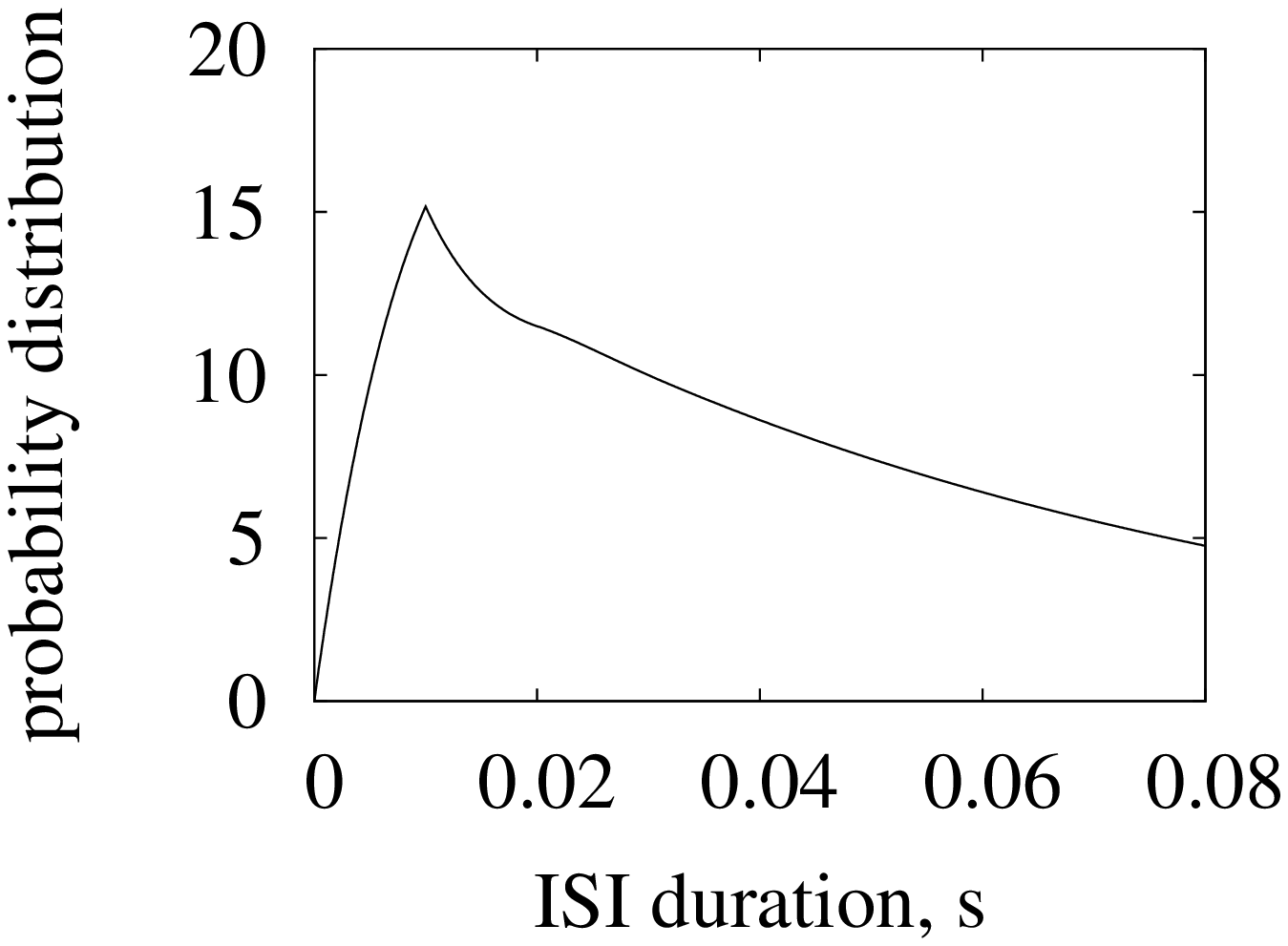}%
	\includegraphics[angle=-90,width=0.45\textwidth]{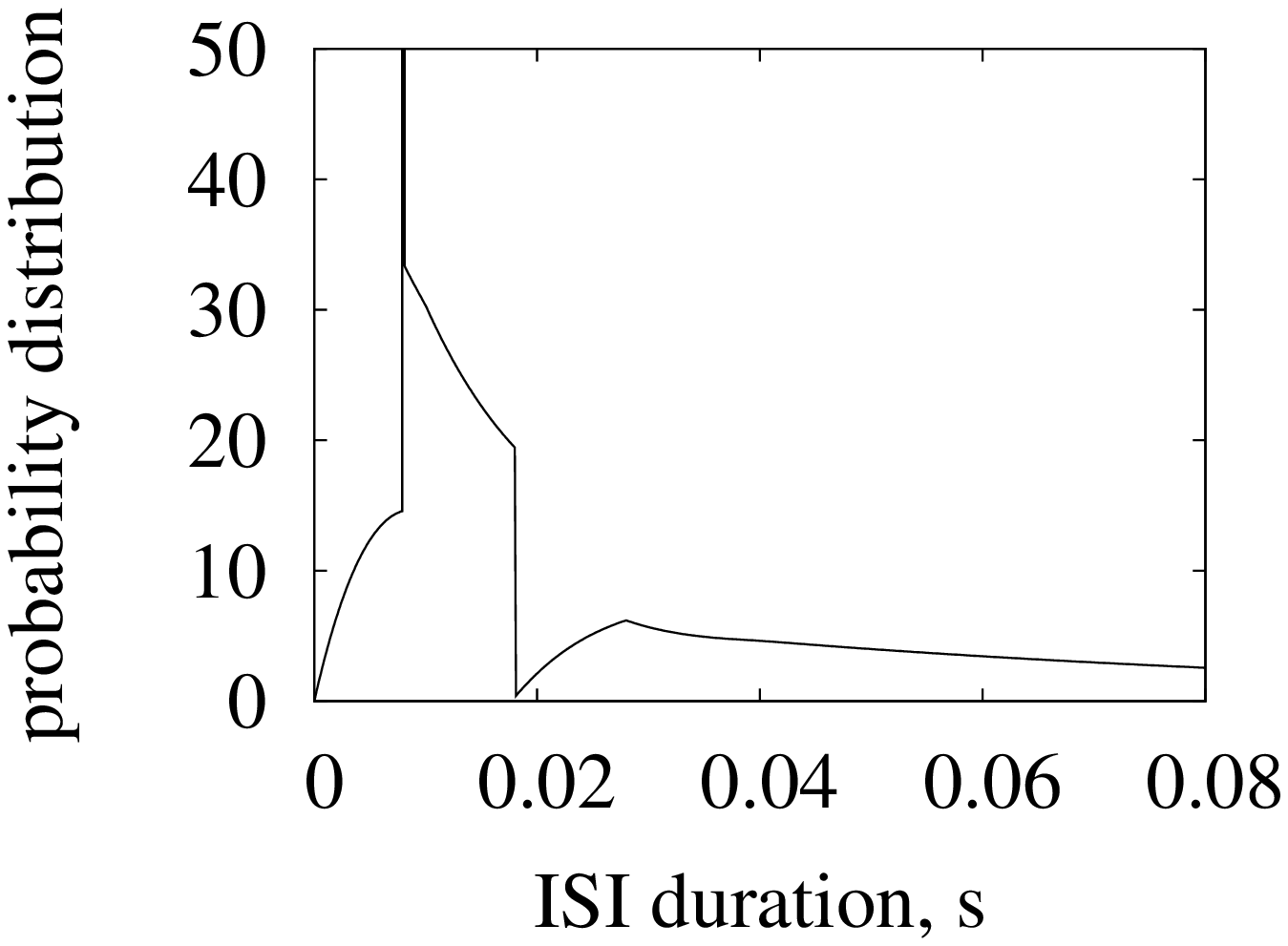}%
\caption{\label{other}ISI probability density $P^{\Delta}(t)$, s$^{-1}$. 
\emph{Left} -- BN without feedback \cite{Vid5};
\emph{right} --
BN with excitatory delayed feedback \cite{BNDF}. 
In both cases, $\tau$ = 10 ms, $\lambda$ = 50 s$^{-1}$, $N_0=2$. For right pannel, $\Delta = 8 $ ms.}
\end{figure}

In order to find $P^\Delta(t)$, one should substitute
(\ref{rozpdr0}) and (\ref{vygliad}) into Eq. (\ref{usered}),
which gives:
\begin{equation}
\label{Pu0}
	P^\Delta(t)= a P^{\Delta}(t\mid\Delta)
	+ \int_{0}^{\Delta} P^{\Delta}(t\mid s)g(s)ds.
\end{equation}
The explicit expression for $P^\Delta(t)$, which can be obtained by further 
transformations of (\ref{Pu0}), will be different for $t$ values belonging to 
different domains. 
This is because the exact expression for the $P^\Delta(t\mid s)$ is different for
different domains, see top and bottom lines of Eq. (\ref{rozpdr0}) and Eq. (\ref{P0}).
The boundaries of domains in which $P^\Delta(t)$ does not change its formula are 
dictated both by the first and the second term in (\ref{Pu0}). Namely, by
taking (\ref{rozpdr0}) with $s=\Delta$
and using (\ref{P0}) one concludes that the first term in 
(\ref{Pu0}) retains the same formula in the adjacent domains separated with points
\begin{equation}\label{boundary1}
t= 0,\, \Delta,\, \tau+\Delta,\, 2\tau+\Delta,\, 3\tau+\Delta,\ldots.
\end{equation}
It appears that the second term in (\ref{Pu0}) as well retains the same formula in
the domain bounded with first two points 
from (\ref{boundary1}) allowing one to obtain exact formula for
the first domain.
Namely, if $t\in]0;\Delta]$, then the first term in (\ref{Pu0}) turns into
\begin{equation}\label{first}
a P^{\Delta}(t\mid\Delta) = a\lambda^2 t\ e^{-\lambda t},
\end{equation}
while integration domain in the second term should be split
 into two parts with the point $s=t$ and use either top, or bottom line of 
 Eq.~(\ref{rozpdr0}) in the corresponding part:
 \begin{multline}\label{second}
 \int_{0}^{\Delta} P^{\Delta}(t\mid s)g(s)ds
 \\=
 \int_{0}^{t} (1+\lambda s)\ e^{-\lambda s} \lambda^2 (t-s) e^{-\lambda (t-s)}\,g(s)\,ds
 \\+
 \int_{t}^{\Delta} \lambda^2 t e^{-\lambda t}\,g(s)\,ds.
 \end{multline}
 By combining Eqs. (\ref{first}) and (\ref{second}), one obtains after transformations
  \begin{multline}\label{t-A}
P^\Delta(t) = \frac{2\lambda\ e^{- \lambda t }}
{2\,\lambda\Delta +3+e^{-2\,\lambda\Delta }} \cdot
\Bigg(\frac{1}{6}\lambda^3 t^3
-\frac{1}{2}\lambda^2 t^2
\\
+\lambda t\Big(\frac{3}{2}+\frac{1}{4}e^{-2\lambda\Delta}
+ \frac{1}{4} e^{-2\lambda(\Delta-t)}\Big)+\lambda^2 t\Delta
\Bigg), \quad t<\Delta.
\end{multline}

When $t\ge\Delta$, one needs only the bottom line of Eq.~(\ref{rozpdr0})
for calculating $P^\Delta(t)$, and Eq.~(\ref{Pu0}) turns into the following:
\begin{multline}\label{tgeD}
	P^\Delta(t)= a\,(1+\lambda\Delta)\ e^{-\lambda \Delta}P^0(t-\Delta) 
\\
+ \int_{0}^{\Delta} (1+\lambda s)\ e^{-\lambda s} P^0(t-s) g(s)\,ds,
\quad t\ge\Delta.
\end{multline}
Here, the first term cannot change its formula within any domain defined
by boundaries (\ref{boundary1}). This is not the case for the second term. 
We analyze the behavior of the second term and obtain explicit expression
for $P^\Delta(t)$ for any $t>0$ in the Appendix. Example graph of $P^\Delta(t)$
found is given at Fig.~\ref{ISI}, {\em left}. Compare with p.d.f. for BN without
feedback, Fig. \ref{other}, left, and BN with excitatory feedback, right.

\subsection{Properties of the ISI probability density \label{prop}}

Notice, that the explicit expressions for $P^\Delta(t)$ given in (\ref{t-A}) and in the
Appendix are not used here. All the properties discussed below are derived from
representation (\ref{usered}) and expressions (\ref{rozpdr0}), 
(\ref{vygliad})--(\ref{a}).

\subsubsection{Mean interspike interval}

%
%
%
%
%
%
\begin{figure}
	\includegraphics[width=0.5\textwidth]{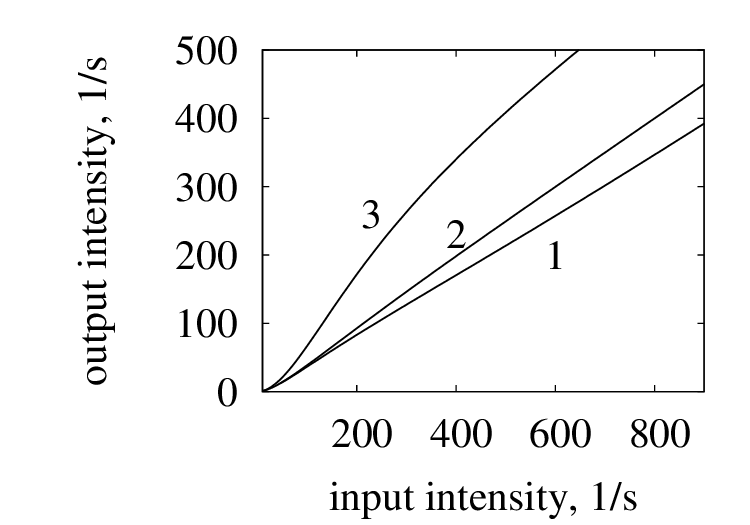}%
	\includegraphics[width=0.5\textwidth]{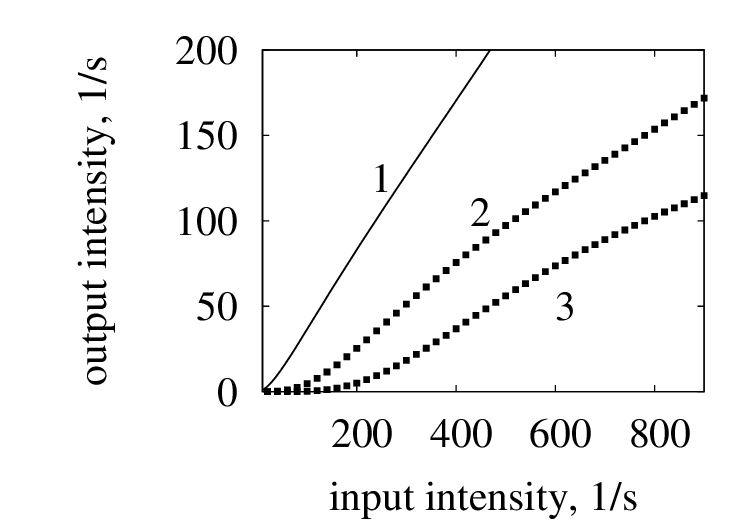}%
\caption{\label{in_out} \emph{Left ---} Mean output firing rate,
$\lambda_o$, {\em vs} 
$\lambda$ for inhibitory BN ($N_{0}=2$) with delayed feedback (\textbf{1}), 
for BN without feedback \cite{BNF} (\textbf{2}) and for excitatory 
BN with
delayed feedback \cite{BNDF} (\textbf{3}), obtained analytically.
\emph{Right ---}
$\lambda_o$ {\em vs}  $\lambda$ for inhibitory 
BN with delayed feedback for $N_0=2$ (\textbf{1}), obtained analytically, and for $N_0=4$ (\textbf{2}) and $N_0=6$ (\textbf{3}), found numerically.  Here $\tau$ = 10 ms for all curves;  $\Delta = 2$ ms for (\textbf{1}), (\textbf{3}), {\em left} 
and (\textbf{1})--(\textbf{3}), {\em right}.}
\end{figure}

The mean output ISI, $W_1^\Delta$, can be defined as the first
moment of the ISI probability density:
\begin{displaymath}
	W_1^{\Delta} = \int_{0}^{\infty} t P^\Delta(t)\,dt.
\end{displaymath}
Taking into account Eq.~(\ref{usered}), one obtains:
$$
	W_1^\Delta = \int_0^\infty t \,dt\,\int_0^\Delta P^\Delta(t\mid s)f(s)\,ds 
	= \int_0^\Delta ds\,f(s)\int_0^\infty t P^\Delta(t\mid s)\,dt,
$$
which taken together with Eq.~(\ref{rozpdr0}) gives:
\begin{equation*}
\begin{split}
	W_1^\Delta &= \int_0^\Delta ds\,f(s)
	\Bigg(
	\int\limits_0^s t^2 e^{-\lambda t}\lambda^2\,dt +
	+(1+\lambda s)\ e^{-\lambda s}	
	\int\limits_{s}^\infty t  P^0(t-s)\,dt\Bigg)
	\\
	&=\frac{1}{\lambda}
	\int\limits_0^\Delta ds\,f(s)
	\left(2 +\, e^{-\lambda s}\left(\lambda W_1^0-2 + (\lambda W_1^0-1)\,\lambda s\right)\,\right),
\end{split}
\end{equation*}
where $W_1^0$ is given in (\ref{W10}).
Use here (\ref{vygliad}) and (\ref{groz}), which gives after transformations:
\begin{equation}
\label{Wd}
	W^\Delta_1= a\,(\Delta + W_1^0),
\end{equation}
where $a$ is given in (\ref{a}).

Note that for $\Delta=0$, Eq. (\ref{Wd}) turns into the following:
\begin{equation}
\nonumber
	W^{\Delta}_1\bigg\vert_{\Delta=0} = W_1^0\,.
\end{equation}
This is consistent with the fact that for $\Delta=0$ 
$P^\Delta(t)$ turns into distribution
for neuron without feedback, which is $P^0(t)$ given in Eqs. (\ref{P0})--(\ref{ym}),
see Eq.~(\ref{PD0}).

The output intensity, $\lambda_{o}$, defined as the mean number of impulses per time unit, is the inverse $W_1^\Delta$:
\begin{equation}
\label{rate}
	\lambda^\Delta_{o} = \frac{1}{W_1^\Delta }= \frac{(2\lambda\Delta+3 + e^{-2\lambda\Delta})(1-e^{-\lambda\tau})}
						{4(\lambda\Delta + 2-(\lambda\Delta +1)
						              e^{-\lambda\tau})}\, \lambda\,,
\end{equation}
where Eqs.~(\ref{W10}), (\ref{a}) and (\ref{Wd}) were used.
At large input rates the following relation takes place
\begin{equation}
\label{lala}
	\lim_{\lambda\to\infty} \left(\frac{\lambda}{2}-\lambda^\Delta_o \right)
	= \frac{1}{4\Delta}.
\end{equation}%
%
%
%
%
\begin{figure}
	\includegraphics[width=0.5\textwidth]{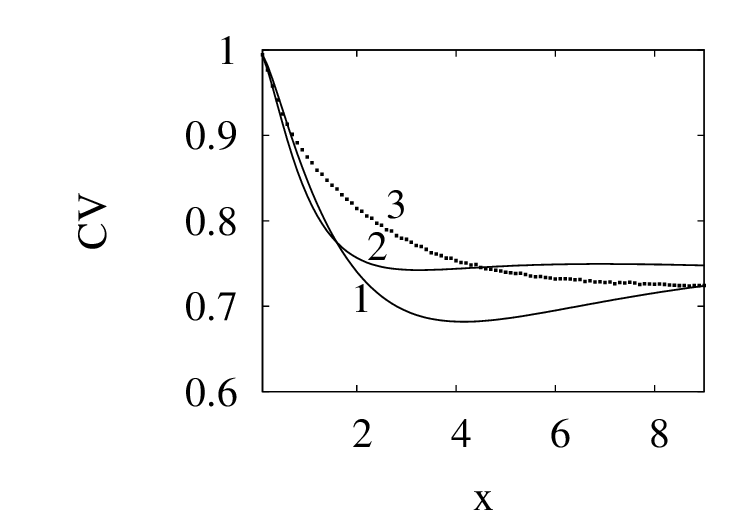}%
	\includegraphics[width=0.5\textwidth]{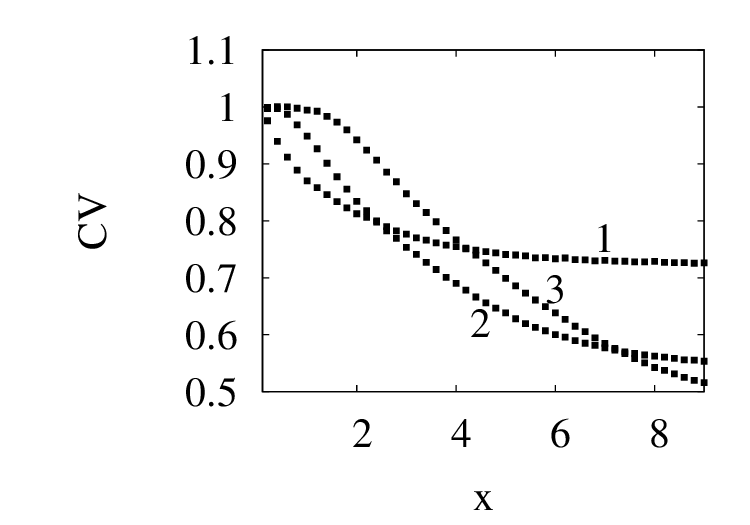}%
\caption{\label{figcv} \emph{Left ---} Coefficient of variation {\em vs} 
$x= \lambda \tau $ for 
BN ($N_{0}=2$)
with inhibitory delayed feedback for $\Delta = 2 $ ms (\textbf{1}), $\Delta = 5 
$ ms (\textbf{2}), obtained analytically, and for $\Delta = 20 $ ms (\textbf{3}),
found numerically.
\emph{Right ---}
Coefficient of variation {\em vs} $x$ for
BN with inhibitory delayed feedback ($\Delta = 18$)
for $N_{0}=2$ (\textbf{1}), $N_{0}=4$ (\textbf{2}) 
and $N_{0}=6$ (\textbf{3}), found numerically.
 $\tau$ = 10 ms for all curves.}
\end{figure}%
This limiting relation, which is derived directly from Eq. (\ref{rate}),
can be understood as follows. At moderate stimulation some input spikes are lost without influencing output due to high probability of long input ISI. 
At high intensity,
every two consecutive excitatory input impulses trigger the BN and send impulse into the feedback line, provided it is empty. Thus, output intensity should be 
$\lambda/2$ minus firings, inhibited by the line. The maximum rate of inhibitory 
impulses, which can be delivered by the feedback line to the neuron's input,
is $1/\Delta$, and this rate is attainable when $\lambda\to\infty$. Each inhibitory impulse
either cancels one excitatory impulse in the neuron, or does nothing if neuron
appears empty at the moment of the feedback line dejection. For high input rates,
the probabilities to find the neuron at any moment of time either
empty, or storing one impulse seem both approaching 0.5.
Thus, due to feedback line activity, about $1/(2\Delta)$ excitatory impulses 
will be eliminated every second from the input stream, and about half as much from the output stream,
which explains (\ref{lala}).

Graphs of $\lambda^\Delta_{o}$ vs $\lambda$ are shown at the Fig.~\ref{in_out}.

\subsubsection{Coefficient of variation}

\begin{figure}
	\includegraphics[width=0.5\textwidth]{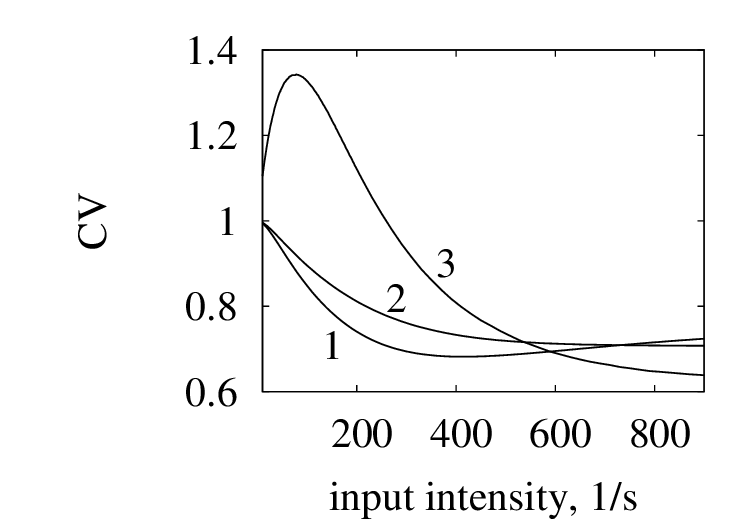}
	\includegraphics[width=0.5\textwidth]{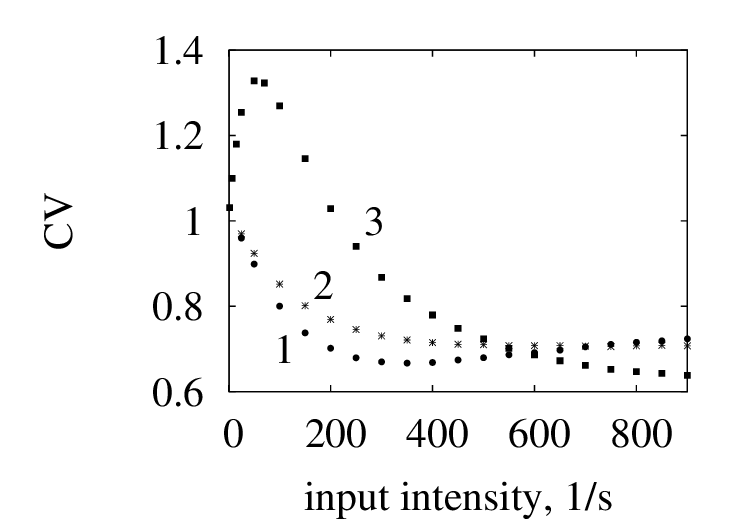}
\caption{\label{cv_feedback} \emph{Left ---} Coefficient of variation {\em vs} 
$\lambda$ for
BN with inhibitory delayed feedback (\textbf{1}), without feedback \cite{BNF} 
(\textbf{2}) and with excitatory delayed feedback \cite{BNDF} (\textbf{3}), 
obtained analytically. Curves (\textbf{1})--(\textbf{3}): $N_{0}=2$, $\tau$ = 10 ms.
\emph{Right ---}
Coefficient of variation {\em vs} $\lambda$ for
LIF neuron with inhibitory delayed feedback (\textbf{1}), 
without feedback (\textbf{2}) excitatory delayed feedback (\textbf{3}), found numerically.
$\Delta = 2$ ms for (\textbf{1}), (\textbf{3}) in both panels.}
\end{figure}

The coefficient of variation (CV) $c_{v}^{\Delta}$ of output ISIs is defined as dimensionless dispersion:
\begin{equation}
\nonumber
	c_{v}^{\Delta} \equiv \sqrt{\frac{W_{2}^{\Delta}}{(W_1^{\Delta})^{2}}-1},
\end{equation}
where $W_{2}^{\Delta}$ is the second moment of the ISI probability density:
\begin{equation}
\nonumber
	W_{2}^{\Delta} \equiv \int_{0}^{\infty}t^{2}\ P^{\Delta}(t)dt
	= \int_{0}^{\Delta}ds\ f(s) \int_{0}^{\infty}t^{2}\ P^{\Delta}(t\mid s)dt.
\end{equation}
By calculating integrals here and taking into account Eq. (\ref{W10}), one obtains:
\begin{equation}
\label{cv}
	(c_{v}^{\Delta})^{2} = 
	\frac{B_{1}\ e^{2 \lambda \tau } + 2\ B_{2}\ e^{ \lambda \tau } + B_{3}}
	{8\Big((2+ \lambda \Delta )\ e^{ \lambda \tau }- \lambda \Delta -1\Big)^{2}}-1,
\end{equation}
where
\begin{equation}
\begin{split}
	B_{1} = & 3\ e^{-4 \lambda \Delta }-8\ e^{-3 \lambda \Delta }+2(6 \lambda \Delta +13)\ e^{-2 \lambda \Delta }-
\\
	& -8(2 \lambda \Delta +3)\ e^{- \lambda \Delta }+12\lambda^{2}\Delta^{2}+52 \lambda \Delta +51,
\\
	B_{2} = & -2\ e^{-4 \lambda \Delta }+4\ e^{-3 \lambda \Delta }
	+2(-5 \lambda \Delta + \lambda \tau -7)\ e^{-2 \lambda \Delta }+
\\
	& +4(2 \lambda \Delta +3)\ e^{- \lambda \Delta } 
	-12\lambda^{2}\Delta^{2}+4\lambda^{2}\Delta\tau-34 \lambda \Delta +6 \lambda \tau -24,
\\
\label{b3}
	B_{3} = & e^{-4 \lambda \Delta }+2(4 \lambda \Delta +3)\, e^{-2 \lambda \Delta }
	+ 12\lambda^{2}\Delta^{2}+24 \lambda \Delta +9
\end{split}
\end{equation}
see Fig.~\ref{figcv}, {\em left}, \ref{cv_feedback}, {\em left}.

It is clear, that for $\Delta=0$ Eqs.~(\ref{cv}), (\ref{b3}) must give the output ISI coefficient of variation of BN without feedback, $c^0_v$. 
And indeed, substituting $\Delta=0$ to (\ref{cv}) and (\ref{b3}), one obtains
\begin{equation}
\nonumber
	c_{v}^2\bigg\vert_{\Delta = 0} = \frac{1}{(2e^{ \lambda \tau }-1)^2}\cdot \Big(2\ e^{2 \lambda \tau }+2( \lambda \tau -1)\ e^{ \lambda \tau }+1\Big)
	=\left(c^0_{v}\right)^2,
\end{equation}
where $c^0_v$ was previously found in \cite[][Sec. 5.3]{BNF}.

\subsubsection{Numerical Simulations}
\label{num}

\begin{figure}
\includegraphics[width=0.45\textwidth]{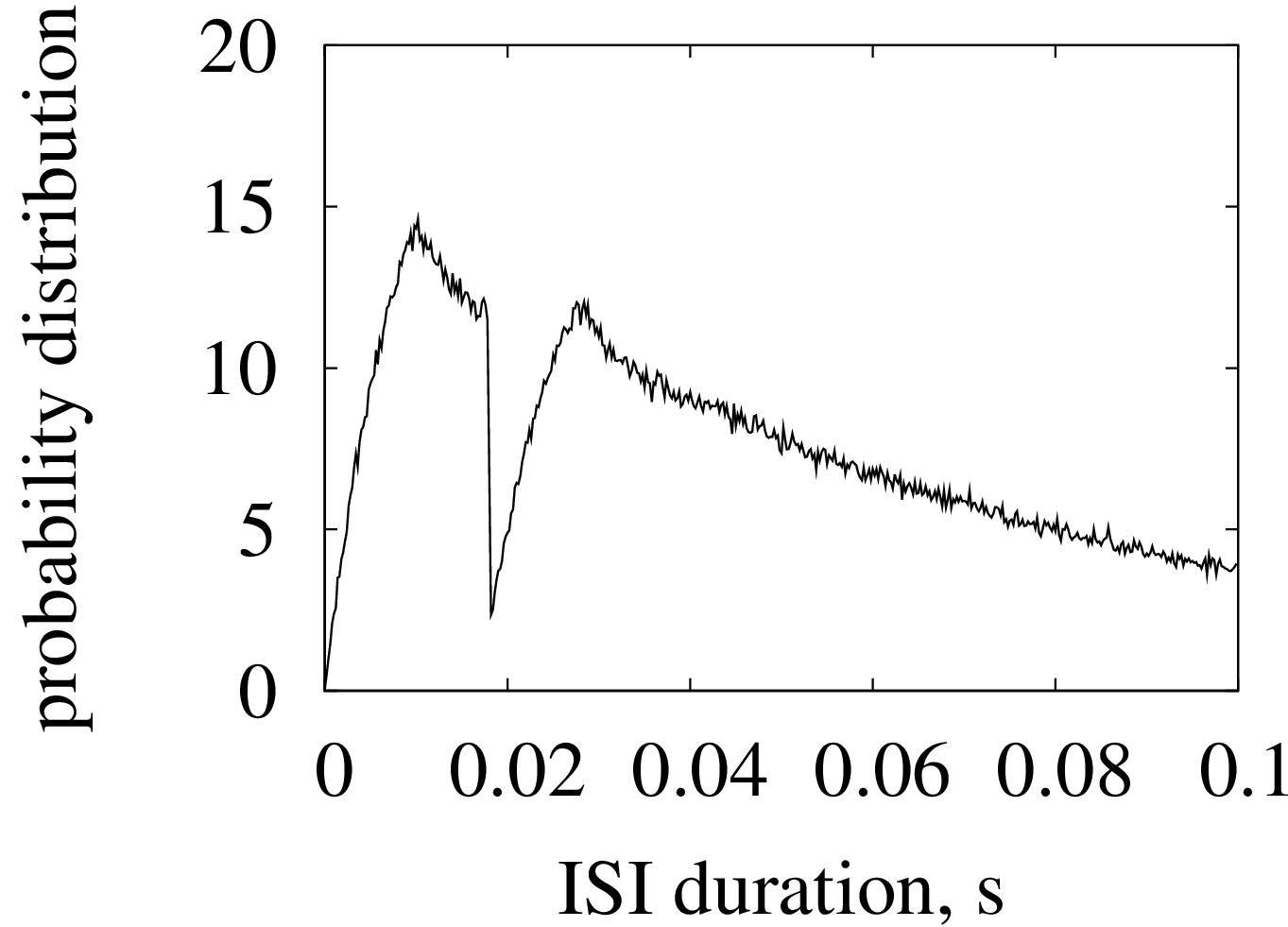}
\includegraphics[width=0.485\textwidth]{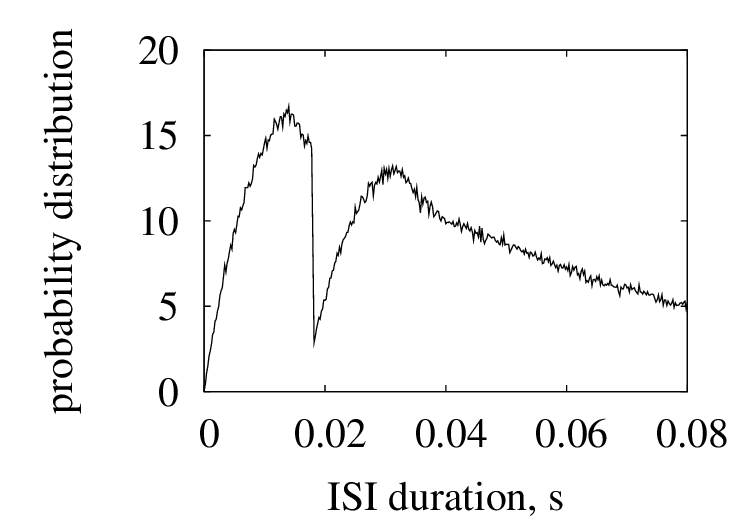}
\caption{\label{LIF} ISI PDF (in s$^{-1}$), found numerically. 
\emph{Left} --- 
inhibitory BN with delayed feedback for $N_{0}=2$, $\tau$ = 10 ms, 
$\lambda = 50$ s$^{-1}$; 
\emph{right} --- inhibitory 
LIF neuron with delayed feedback. $\Delta = 18 $ ms in both panels.}
\end{figure}

The numerical simulations were performed as described in Sec. \ref{MeNum}.
BN firing statistics is represented 
in terms of $P^{\Delta}(t)$, $f(s)$, $\lambda_{o}^{\Delta}$ and $c^{\Delta}_v$.
In parallel to the analytic expressions, all these quantities were calculated 
numerically for various sets of parameters $\tau$, $\Delta$, $\lambda$.
Numerically calculated curves were then compared with corresponding 
analytic expressions
given in Eqs.~(\ref{t-A}), (\ref{eq:P1}) -- (\ref{eq:rho}),
(\ref{vygliad}) -- (\ref{a}), 
(\ref{rate}),
(\ref{cv}) and (\ref{b3}).
It was found, that numerically obtained curves fit perfectly with mentioned 
analytic expressions, see example in Fig.~\ref{ISI}.

\begin{figure}
	\includegraphics[width=0.35\textwidth,angle=-90]{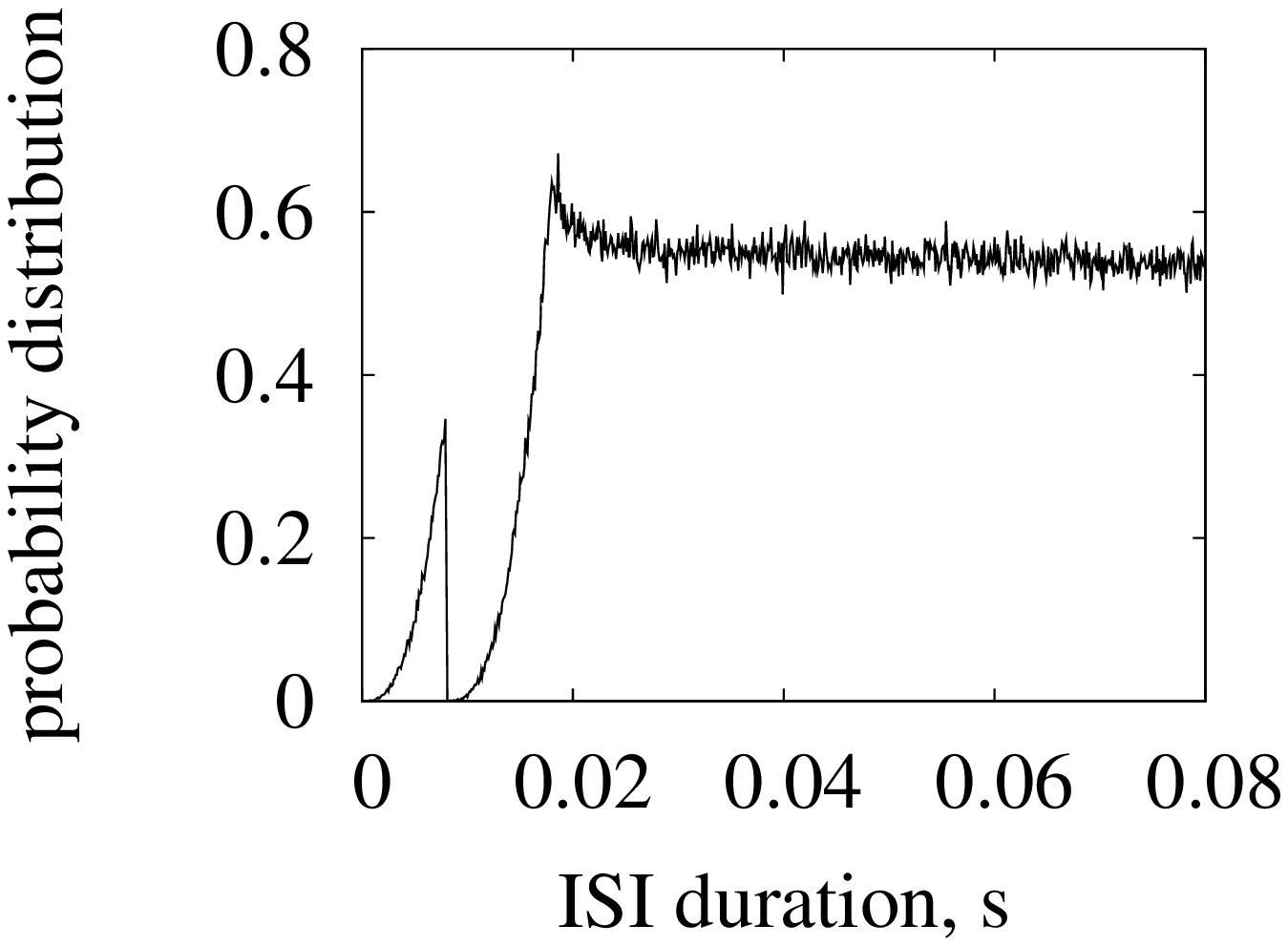}
	\includegraphics[width=0.35\textwidth,angle=-90]{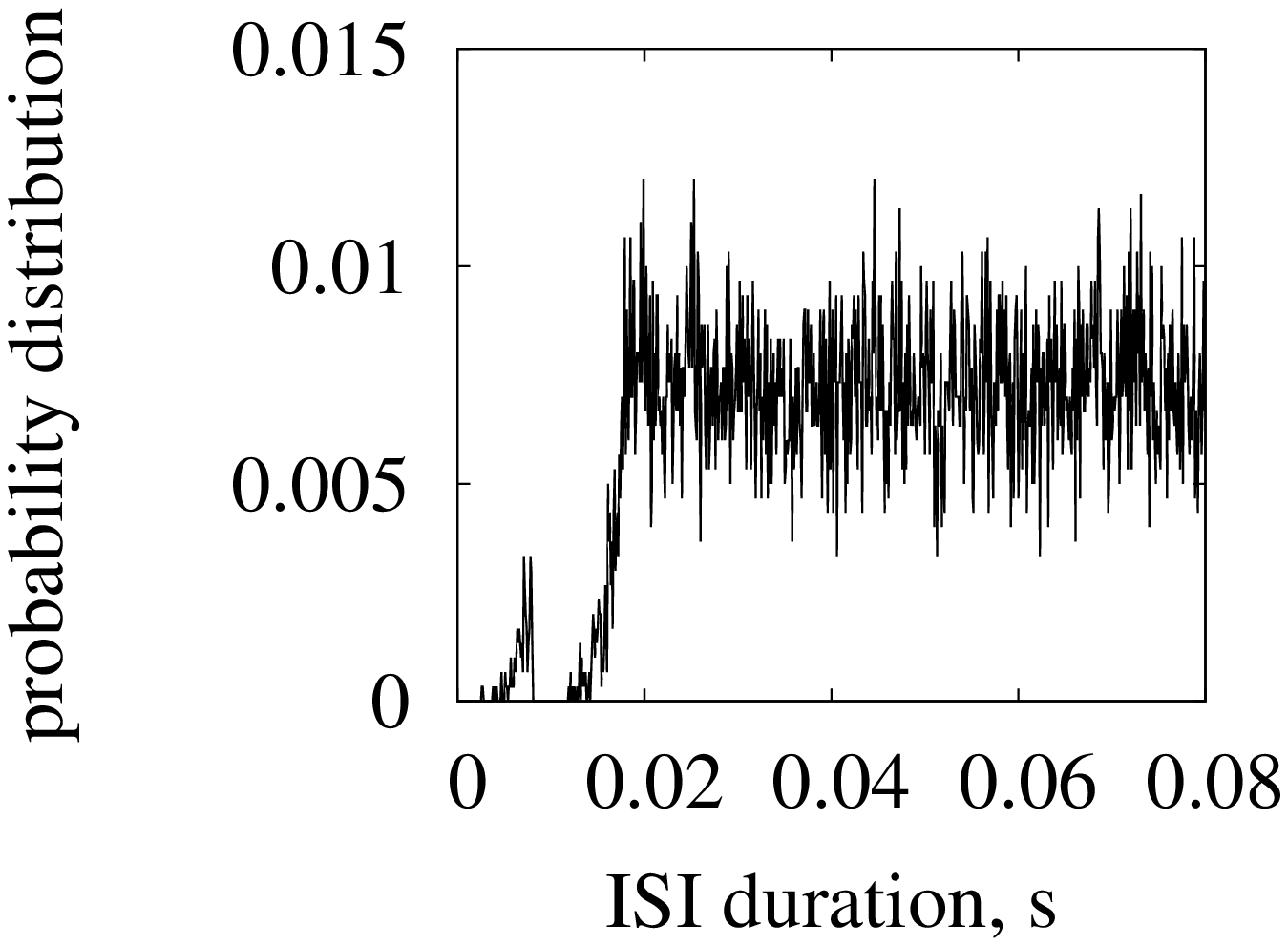}
\caption{\label{Th46}ISI probability density $P^{\Delta}(t)$ (measured in s$^{-1}$) found numerically for $\tau$ = 10 ms, $\Delta = 8 $ ms, $\lambda$ = 50 s$^{-1}$. \emph{Left} --- $N_0=4$, \emph{right} --- $N_0=6$.  In both cases $3\cdot10^7$ triggerings were taken.}
\end{figure}

The set of numerical simulations was performed for the case $\Delta > \tau$
and/or $N_0>2$, which is not covered by the analytic expressions obtained.
The curves obtained are given in Fig. \ref{in_out}, {\em right}, (\textbf{2}) and
(\textbf{3}), Fig. \ref{figcv}, {\em left}, (\textbf{3}), {\em right}, 
(\textbf{1})-(\textbf{3}), Fig. \ref{LIF}, {\em left}, Fig. \ref{Th46}.

A number of numerical simulations were performed for the LIF neuron model
with parameters given in Sec. \ref{MeNum}. The data obtained is presented
in Fig. \ref{cv_feedback}, {\em right}, Fig. \ref{LIF}, {\em right}.

\section{Discussion}

The statement of the problem adopted in this work expects transformation
of the input ISI PDF into the output ISI PDF. 
In the contrast to widely used diffusion approximation, 
which lacks the input ISI PDF, we treat
both input and output streams as objects of the same mathematical nature,
namely, the point stochastic processes. This is necessary
if one expects to study activity of a neuron involved in the
interneuronal communication with the time-coding 
as opposed to the rate-coding paradigm.
This approach allowed us to consider a system, where output impulses
are conveyed directly to its input --- neuron with feedback --- the simplest
case of interneuronal communication. The feedback is inhibitory here,
which is in concordance with numerous self-inhibitory neurons
observed in nature, see \cite{Bacci,Nicoll,Smith}. As a result, we 
obtained exact expressions for several quantities characterizing the activity
of binding neuron with feedback, compared those quantities with
those calculated numerically for the LIF neuron and made the following
conclusions.

\subsection{Conclusions}

In this paper the binding neuron model with delayed self-inhibi\-ti\-on 
is studied. The neuron is stimulated with point stochastic process --- Poisson 
stream of given intensity. The ISI PDF was found as an exact function of the 
input intensity, $\lambda$, delay time, $\Delta$ and neuronal internal memory 
duration time, $\tau$ for the BN neuron with threshold $N_0=2$.  
The ISI PDF for higher thresholds is found numerically.
The course of the PDFs found is bimodal due to a trough at $t=\Delta$,
see Figs.~\ref{ISI}, \ref{LIF}, \ref{Th46}. The nature of the trough is the
same as is the nature of peak in the case of excitatory feedback, which was 
discussed by  L.Ricciardi, \cite{Ricciardi}, and exactly calculated for the BN in
\cite{BNDF}.

 Exact mathematical expressions are found for the mean interspike interval,
and coefficient of variation. Those same quantities are found numerically for
the leaky integrate and fire model with shunting type delayed self-inhibition.
Both models studied deliver qualitatively similar results. We conclude that
the character of dependencies found is mainly due to the presence of
inhibitory feedback line.


\appendix
\section{}

Let us introduce a new variable of integration, $u=t-s$ in the Eq. (\ref{tgeD}):
\begin{multline}\label{t-Du}
	P^\Delta(t)= a\,(1+\lambda\Delta)\ e^{-\lambda \Delta}P^0(t-\Delta) 
\\
	+ \int_{t-\Delta}^{t} (1+\lambda(t-u))e^{-\lambda (t-u)} P^0(u) g(t-u)\,du.
\end{multline}
Here, due to (\ref{P0})--(\ref{ym}),
both the first term and the second one can change its formula with changing $t$
value. The first term changes its form every time when
$t-\Delta$ crosses integer multiple of $\tau$. This gives the boundary points
(\ref{boundary1}). The second term in (\ref{t-Du}) can change its value every
time when either $t$, or $t-\Delta$ crosses integer multiple of $\tau$. This gives
the following boundary points
\begin{equation}\label{boundary2}
t= 0,\, \Delta,\, \tau,\, \tau+\Delta,\, 2\tau,\, 2\tau+\Delta,\, 3\tau,\, 3\tau+\Delta,\ldots.
\end{equation}
If the $t$ value varies between two consecutive points from (\ref{boundary2}),
then both terms in (\ref{tgeD}) or (\ref{t-Du}) retain their algebraic expression.
In order to find explicit form of that expression at any domain of $t$ defined by 
(\ref{boundary2}), we introduce two groups of domains, $B_m$ and $C_m$, namely:
\begin{align}\nonumber
	B_m =~ & [m\tau+\Delta;(m+1)\tau]\,, &m=0,1,\ldots,
\\\nonumber
	C_m =~ & ](m+1)\tau;(m+1)\tau+\Delta[\,, &m=0,1,\ldots.
\end{align}
Note, that the full range $]0;\infty[$ of possible ISI values is covered by
alternate domains $B_m$ and $C_m$, $m=0,1,\ldots$ and the domain 
$]0;\Delta[$ for 
which we already have the explicit formula (\ref{t-A}). 

If $t\in B_m$, then $m\tau\le t-\Delta < t\le(m+1)\tau$, and one should substitute 
$y_{m}(t)$ from (\ref{ym}), corresponding to that $m$, instead of $P^0(u)$ in the (\ref{t-Du}). If $t\in C_m$, then $m\tau< t-\Delta < (m+1)\tau < t$. Therefore, the domain of integration in the Eq. (\ref{t-Du}) should be split into two with point $(m+1)\tau$, and as $P^0(u)$ one should substitute either $y_{m}(t)$, or $y_{m+1}(t)$.

\subsection{ISI probability density at the domains $B_m$}

Thus, in the case $t\in B_m$, one obtains for $P^{\Delta}(t)$:
\begin{multline}
	P^{\Delta}(t) = a\,(1+ \lambda \Delta )\ e^{-\lambda \Delta}y_{m}(t-\Delta)
\\
\label{Pbm}
	+\int_{0}^{\Delta}  (1+\lambda s)\ e^{-\lambda s} y_{m}(t-s)g(s)ds,
\quad t\in B_m,
\end{multline}
which after integration gives:
\begin{multline}
\label{eq:P1}
P^{\Delta}(t)\ =\ a\,(1+\lambda\Delta)\ e^{-\lambda\Delta}\cdot y_m(t-\Delta)
-\frac{a}{2}\ e^{\lambda(\Delta-\tau)}\cdot y_{m+1}(t-\Delta+\tau) 
\allowdisplaybreaks\\
	+\frac{a}{2}\ e^{ \lambda \tau }\cdot y_{m+1}(t+\tau)
	+\frac{a\lambda}{2}\ e^{- \lambda t }
	\sum\limits_{k=1}^{m+1}\sum\limits_{l=0}^{k}
	K_{kl}\lambda^{k-l}(t-(k-1)\tau)^{k-l}
\allowdisplaybreaks\\
	-\frac{a\lambda}{2}\ e^{- \lambda t }
	\sum\limits_{k=1}^{m}\sum\limits_{l=0}^{k}
	K_{kl} \lambda^{k-l}(t-k\tau)^{k-l},
	\quad t\in B_m,
\end{multline}
where
\begin{multline}\nonumber
	K_{kl} = \frac{1}{2^{l+2}(k-l)!} 
	\bigg(\frac{l+1}{(l+2)!}\,(-2 \lambda \Delta )^{l+2} + l+1
\allowdisplaybreaks\\
	-(l-1)\,e^{-2 \lambda \Delta }
	-2 \sum_{i=0}^{l}
	\frac{(-2 \lambda \Delta )^{l-i}}{(l-i)!}
	\left(1+\frac{l+1}{l+1-i}\cdot \lambda \Delta \right)\bigg). 
\end{multline}

\subsection{ISI probability density at the domains $C_m$}

Consider the case $t\in C_m$. Taking into account Eqs. (\ref{P0})--(\ref{ym}), 
one can rewrite (\ref{t-Du}) as follows 
\begin{multline}\nonumber
P^{\Delta}(t)\bigg\vert_{t\in C_m}  \hfill
\\	\allowdisplaybreaks
=a\,(1+ \lambda \Delta )\ e^{-\lambda \Delta}y_{m}(t-\Delta)
+\int_{0}^{t-(m+1)\tau} (1+\lambda s)\ e^{-\lambda s} y_{m+1}(t-s)g(s)ds
\hfill
\\	\allowdisplaybreaks
+\int_{t-(m+1)\tau}^{\Delta} (1+\lambda s)\ e^{-\lambda s} y_{m}(t-s)g(s)ds 
\\
	\,=\,
a\,(1+ \lambda \Delta )\ e^{-\lambda \Delta}y_{m}(t-\Delta)
	+\int_{0}^{\Delta}  (1+\lambda s)\ e^{-\lambda s} y_{m}(t-s)g(s)ds\hfill
\\\allowdisplaybreaks
	+\frac{\lambda^{m+3}}{(m+2)!}\ e^{- \lambda t } 
	\int_{0}^{t-(m+1)\tau} (1+\lambda s)\,(t-s-(m+1)\tau)^{m+2} g(s)ds
\\\allowdisplaybreaks
	-\frac{\lambda^{m+2}}{(m+1)!}\ e^{- \lambda t }
	\int_{0}^{t-(m+1)\tau} (1+\lambda s)\,(t-s-(m+1)\tau)^{m+1} g(s)ds.
\end{multline}
It is useful to denote as $P_{B,m}^\Delta(t)$ the right-hand side of Eq. (\ref{Pbm})
defined for all $t$: 
\begin{multline}\nonumber
	P_{B,m}^\Delta(t) = a\,(1+ \lambda \Delta )\ e^{-\lambda \Delta}y_{m}(t-\Delta)+
\int_{0}^{\Delta}  (1+\lambda s)\ e^{-\lambda s} y_{m}(t-s)g(s)ds,
\quad t>0.
\end{multline}
With this notation, one obtains:
\begin{equation}
\label{eq:P2}
	P^{\Delta}(t)\bigg\vert_{t\in C_m}=
	P_{B,m}^\Delta(t)
	+\frac{a\lambda}{2}\ e^{- \lambda t }\cdot\rho^{\Delta}_{m}\left(\lambda\left(t-(m+1)\tau\right)\right),
\end{equation}
where
\begin{multline}
\label{eq:rhodef}
	\rho^{\Delta}_{m}(x)= 
	\frac{2}{a\lambda} \frac{1}{(m+2)!}
	\int\limits_{0}^{x} (1+v)\,(x-v)^{m+2} g\left(\frac{v}{\lambda}\right)dv
\\
	-\frac{2}{a\lambda}\frac{1}{(m+1)!}
	\int\limits_{0}^{x} (1+v)\,(x-v)^{m+1} g\left(\frac{v}{\lambda}\right)dv,
\quad m=0,1,\ldots.
\end{multline}
Here the dimensionless variable of integration $v=\lambda s$ was introduced.

Performing integration in (\ref{eq:rhodef}) one obtains:
\begin{equation}
	\rho^{\Delta}_{m}(x)= 
	\sum\limits_{l=0}^{m+4}K_l\ x^{l}
	-e^{-2 \lambda \Delta +2x}\cdot \sum\limits_{l=0}^{m+3}D_l\ x^{l},
\end{equation}
where
\vspace{-0.5\baselineskip}
\begin{gather}\label{eq:rho}
\begin{split}
D_l &=\frac{1}{2^{m+4-l}}\cdot\sum\limits_{i=0}^{l}\frac{(-1)^{l-i}\cdot (m-1-i)}{i!\ (l-i)!}, 
\\\\\allowdisplaybreaks
K_l &=\frac{(m-1-l)}{2^{m+4-l}\cdot l!}\ e^{-2 \lambda \Delta },\quad l=0,\ldots m+1, 
\allowdisplaybreaks\\ \text{ }
\allowdisplaybreaks\\
K_{m+2} &=\frac{1}{4\cdot (m+2)!}\ e^{-2 \lambda \Delta }-\frac{1}{(m+2)!},
\allowdisplaybreaks\\
K_{m+3} &=\sum\limits_{i=1}^{m+2}\frac{(-1)^i\ i}{(m+2-i)!\ (i+1)!}+\frac{1}{(m+3)!}, 
\allowdisplaybreaks\\
K_{m+4} &= \sum\limits_{i=0}^{m+2}\frac{(-1)^i\ (i+1)}{(m+2-i)!\ (i+2)!}, 
\allowdisplaybreaks\\
D_{m+2} &=\frac{1}{4}\sum\limits_{i=0}^{m+2}\frac{(-1)^i\cdot(i+1)}{i!\ (m+2-i)!},\qquad 
D_{m+3} =\frac{1}{2}\sum\limits_{i=0}^{m+2}\frac{(-1)^i}{i!\ (m+2-i)!}\,.
\end{split}
\end{gather}

Note, that in the case $\Delta=0$, ISI probability density is completely defined by Eq. (\ref{eq:P1}), which turns into probability distribution for BN without feedback
given in (\ref{P0}):
\begin{equation}
\label{PD0}
	P^{\Delta}(t)\vert_{\Delta=0}= P^{0}(t),
	\quad t>0.
\end{equation}
This indeed should be the case, because when $\Delta=0$, inhibitory impulses always enter empty neuron and, therefore, the feedback line have no chance to affect the output stream.
Naturally, the output ISI distribution for $\Delta=0$ coincides with that found for BN without feedback.


\begin{thebibliography}{00}

\bibitem{Aron}
V.~Aroniadou-Anderjaska, M.~Ennis and M.T.~Shipley,
Dendrodendritic recurrent excitation in mitral cells of the rat
olfactory bulb,
J. Neurophysiol.  {82}  (1999)  489--494.

\bibitem{Bacci}
A.~Bacci, J.R.~Huguenard and D.A.~Prince,
Functional autaptic neurotransmission in fast-spiking interneurons:
A novel form of feedback inhibition in the neocortex,
J. Neurosci.  {23}  (2003)  859--866.

\bibitem{Bacci2004}
A.~Bacci, J.R.~Huguenard and D.A.~Prince,
Long-lasting self-inhibition of
neocortical interneurons mediated by
endocannabinoids,
Nature  {431}  (2004)  312--316.

\bibitem{Adeli}
S. Ghosh-Dastidar, H. Adeli,
Spiking neural networks,
Intern. J. Neural Sys. {19} (2009) 295--308.

\bibitem{Gulyas}
Gulyas,A.I., Miles,R., S\'{\i}k,A., T\'oth,K., Tamamaki,N., Freund,T.F.
Hippocampal pyramidal cells excite inhibitory neurons through a single release site
Nature 336(6456):683-687 (1993)

\bibitem{Jonas}
Jonas,P., Bischofberger,J., Fricker,D., Miles,R.
Fast in, fast out ­ temporal and spatial signal processing in hippocampal interneurons
Trends in Neurosciences 27(1):30-40 (2004)


\bibitem{Kostur}
M. Kostur, P. H\"anggi, P. Talkner, and J. L. Mateos
Anticipated synchronization in coupled inertial ratchets with time-delayed feedback: A numerical study,
Phys. Rev. E 72 (2005) 036210.

\bibitem{MacKay} 
D.M.~MacKay, 
Self-organization in the time domain, 
in: M.C.~Yovitts, G.T.~Jacobi and G.D.~Goldstein (Eds.),
{Self-Organizing Systems},   Spartan Books, Washington, 1962, pp.\,37--48.

\bibitem{Nawrot2007}
M.P.~Nawrot, C.~Boucsein, V.~Rodriguez-Molina, A.~Aertsen, S.~Gr\"un and S.~Rotter,
Serial interval statistics of spontaneous activity in cortical neurons in vivo and in vitro,
Neurocomputing  {70}  (2007)  1717--1722.

\bibitem{Nicoll}
R.A.~Nicoll and C.E.~Jahr,
Self-excitation of olfactory bulb neurons,
Nature  {296}  (1982)  441--444.

\bibitem{Ricciardi}
L.M.~Ricciardi,
Analysis of spike trains in terms of delayed coincidences,
Experimental Brain Research  {3}  (1967)  1-11.


\bibitem{Sakmann}
T.W.~Margrie, B.~Sakmann and N.N.~Urban,
Action potential propagation in mitral cell lateral
dendrites is decremental and controls recurrent and lateral inhibition
in the mammalian olfactory bulb,
PNAS  {98}   (2001)  319--324.

\bibitem{Miles}
Miles,R.
Synaptic excitation of inhibitory cells by single CA3 hippocampal pyramidal cells of the guinea-pig in vitro
Journal of Physiology 428:61­-­77 (1990)



\bibitem{Sch}
R.F.~Schmidt, Fundamentals of Neurophysiology, Springer, 1981.

\bibitem{Segundo1963}
J.P.~Segundo, G.P.~Moore, L.J.~Stensaas and T.H.~Bullock,
Sensitivity of neurons in Aplysia to temporal pattern of arriving impulses,
J.exp.Biol.  {40}  (1963)  643--667.

\bibitem{Segundo} 
J.P.~Segundo, D.~Perkel, H.~Wyman, H.~Hegstad and G.P.~Moore, 
Input-output relations in computer-simulated nerve cell, 
Kybernetic  {4}  (1968)  157--171.

\bibitem{Smith}
T.C.~Smith and C.E.~Jahr,
Self-inhibition of olfactory bulb neurons,
Nature Neuroscience  {5}  8, (2002)  760--766.

\bibitem{Softky} 
W.R.~Softky and C.~Koch, 
The highly irregular firing of cortical cells is inconsistent with temporal integration of random EPSPs, 
J. Neurosci.  {13}  (1993)  334--350.

\bibitem{Vid} 
A.K.~Vidybida, 
Neuron as time coherence discriminator, 
Biol. Cybern.  {74}  (1996)  539--544.

\bibitem{Vid3} 
A.K.~Vidybida, 
Inhibition as binding controller at the single neuron level, 
BioSystems  {48}  (1998)  263--267.

\bibitem{Vid5} 
A.K.~Vidybida, 
Output stream of a binding neuron, 
Ukrainian Mathematical Journal  {50}  (2007)  1819--1839.

\bibitem{BNF} 
A.K.~Vidybida, 
Output stream of binding neuron with instantaneous feedback, 
Eur. Phys. J. B  {65}  (2008)  577--584;
Eur. Phys. J. B  {69}  (2009) 313.

\bibitem{BNDF} 
A.K.~Vidybida, K.G.~Kravchuk, 
Output stream of binding neuron with delayed feedback, 
Eur. Phys. J. B  {72}  (2009)  279--287.

\bibitem{Zhang}
Pan Zhang, Yong Chen ,
Topology and dynamics of attractor neural networks: The role of loopiness,
Physica A 387 (2008) 4411­--4416.





\end{thebibliography}
\end{document}